\documentclass[preprint,showpacs,amsmath,amssymb]{revtex4}

\usepackage{graphicx} 
\usepackage{epsfig}

\begin{document}

\title{From short to fat tails in financial markets: A unified description}

\author{A.A.G. Cortines} 
\affiliation{  Departamento de F\'{\i}sica, Pontif\'{\i}cia 
Universidade Cat\'olica do Rio de Janeiro,  
CP 38097, 22453-900, Rio de Janeiro, Brazil                  
}
\author{R. Riera} 
\affiliation{  Departamento de F\'{\i}sica, Pontif\'{\i}cia 
Universidade Cat\'olica do Rio de Janeiro,  
CP 38097, 22453-900, Rio de Janeiro, Brazil                  
}
\author{C. Anteneodo}
\affiliation{  Departamento de F\'{\i}sica, Pontif\'{\i}cia 
Universidade Cat\'olica do Rio de Janeiro,  
CP 38097, 22453-900, Rio de Janeiro, Brazil                  
}

\email{cortines@fis.puc-rio.br}
\email{rrif@fis.puc-rio.br}
\email{celia@fis.puc-rio.br}

\begin{abstract}
In complex systems such as turbulent flows and financial markets, 
the dynamics in long and short time-lags,   
signaled by Gaussian and fat-tailed statistics,  
respectively, calls for a unified description. 
To address this issue we analyze a real dataset, 
namely, price fluctuations, in a wide range of temporal scales to 
embrace both regimes.
By means of Kramers-Moyal (KM) coefficients evaluated 
from empirical time series, we obtain the evolution equation for the 
probability density function (PDF) of price returns.  
We also present consistent asymptotic solutions for the timescale dependent 
equation that emerges from the empirical analysis.
From these solutions, new relationships connecting PDF characteristics, such as tail exponents, 
to parameters of KM coefficients arise.
The results reveal a dynamical path that leads from Gaussian to fat-tailed 
statistics, furnishing insights on other complex systems where 
akin crossover is observed. 
\end{abstract}


\maketitle

One of the main problems in statistical physics consists 
in the study of macroscopic changes of systems in which 
fluctuations play a central role, 
e.g., diffusion and noise-induced transitions.
The Fokker-Planck equation (FPE) \cite{risken} provides a powerful tool 
for dealing with such problems and has been used in many different fields 
in natural sciences, including solid-state and plasma physics, quantum optics, 
chemical and nuclear reaction kinetics, molecular biology and population dynamics.
Financial data have also been described as stochastic processes 
governed by Langevin or FPEs. These efforts are of utter relevance due to 
the strongly complex fluctuating dynamics of financial time series, 
which poses new challenges to model the dynamical laws responsible 
for the observed statistical properties. 
In that direction, for example,  the anomalous diffusive properties 
and second-order correlations of price fluctuations (e.g., see \cite{book}), 
have been addressed through multivariate \cite{yakovenko} and non-linear \cite{lisa} 
models.
However, these attempts are usually built phenomenologically and, as far
as they are manifold, entail drawbacks for unmasking the processes 
that rule the underlying dynamics.

A typical feature of complex systems is the existence of non-trivial structures 
on different timescales. In particular, the dynamics of price fluctuations in 
long and short timescales, signaled by Gaussian and fat-tailed probability density 
functions (PDFs), 
respectively, has been usually treated in the literature separately, and a unified 
description in the full range of timescales is still lacking. 

Our present goal is two-fold.  The first one is to grasp, through a non-parametric 
method, the underlying stochastic dynamics of price fluctuations 
in order to unveil the driving mechanisms within a unified framework. 
The second one is to determine analytically, from the 
stochastic equations that emerge from the first step, the family of 
PDFs that encompasses the observed timescale dependent ones.
To these ends, we obtain the evolution equation for the PDF of price 
returns through the estimation of Kramers-Moyal (KM) 
expansion coefficients. 
We follow the work by Friedrich and collaborators \cite{friedrich,turbulence}, 
which exploits a correspondence between financial market dynamics 
and hydrodynamic turbulence \cite{turbfinance} assuming the 
existence of a flux of information towards finer scales.  
This approach has been applied before for developed markets, 
although for limited time windows \cite{friedrich,amjap,ausloos,peinke}.
Here, as a relevant example, we scrutinize  the daily and intraday time 
series of Ibovespa, the financial index 
of the Brazilian stock market, which is not fully understood,  
despite typifying major emergent markets.  
We address a large hierarchy of 
time-lags, ranging from months to minutes.
From the measured KM coefficients, we are able to reproduce the full evolution 
of empirical histograms of price returns, embracing the crossover 
from Gaussian to strongly fat-tailed PDFs when going from 
large to short timescales. 
We also present consistent solutions of the resulting FPE. 
They belong to the class of generalized Student-t distributions 
(also known in recent literature as 
$q$-Gaussians  \cite{qgaussian}) that comprises the non-stationary invariant 
PDFs  observed in both asymptotic limits.

We investigate the timescale evolution of the PDF of logarithmic price 
increments (returns) $\Delta x$. 
We consider, as a general evolution equation for those PDFs,  
the KM expansion of a master equation, valid for Markov 
processes \cite{risken}: 
\begin{equation} \label{KM}
\frac{\partial P(\Delta x,\tau)}{\partial\tau}  = \sum_{k\geq1}
\Bigl[-\frac{\partial}{\partial \Delta x}\Bigr]^k D^{(k)}(\Delta x,\tau) P(\Delta x,\tau),  
\end{equation}
where the coefficients  $D^{(k)}(\Delta x,\tau)$ are defined as 
\begin{equation} \label{Dk}
D^{(k)}(\Delta x,\tau) = \lim_{\Delta \tau \to 0} 
\tilde D^{(k)}(\Delta x,\tau,\Delta\tau) ,
\end{equation}
with  $\tilde D^{(k)}=
 M^{(k)}(\Delta x, \tau,\Delta\tau)/\Delta \tau/k!$,
being $M^{(k)}$ the moments of the conditional PDFs, i.e.,
$ M^{(k)}= \int d \Delta x^\prime  
(\Delta x^\prime - \Delta x)^{k} P(\Delta x^\prime, 
\tau+\Delta\tau| \Delta x, \tau)$.  
Following the insight provided by cascade models in turbulence, 
we consider a logarithmic time scale defined as $\tau =\log_2(\Delta t_0/\Delta t)$, 
where $\Delta t_0$ is a reference time-lag.

For the analysis of Ibovespa, we select three datasets: 3960 deflated daily 
closing prices, in the term 02Jan.1991-28Dec.2006, 
37984 15-minute cotes in 21Jan.1998-31Mar.2003 and 794310 30-second 
cotes in 01Nov.2002-19Jul.2006. 
The timescale $\Delta t_0$ ($\tau=0$) is set as 32 (trading) days. 
In what follows, the measured returns are given in 
units of the standard deviation $\sigma_{32}$ of the respective data sample 
at 32-day time-lag. 

\begin{figure}[t!]
\centering
\includegraphics*[bb=80 458 550 724, width=0.7\textwidth]{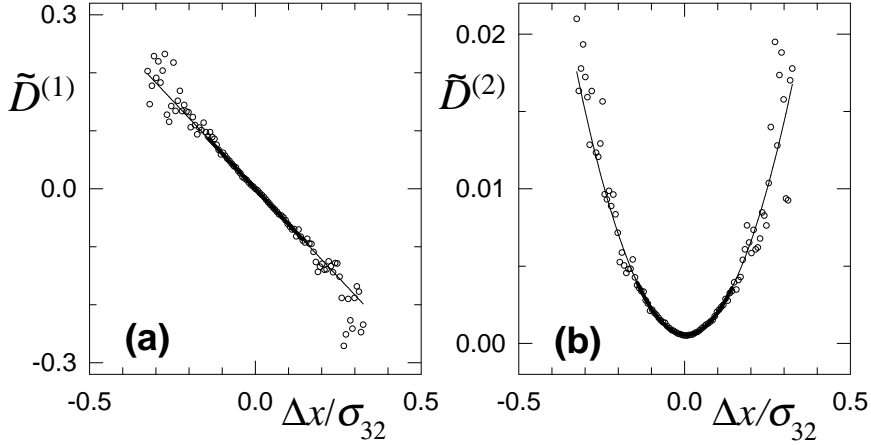}  
\caption{Coefficients $\tilde D^{(1)}$ (a) and 
$\tilde D^{(2)}$ (b) as a function of $\Delta x$, evaluated at  
$\Delta t_2=109$ min and $\Delta t_1=125$ min (hence $\tau=7.9$ and $\Delta\tau=0.2$). 
The abscissa axes are rescaled by the standard 
deviation $\sigma_{32}$ of the returns at 32-day time-lag. 
Solid lines correspond to the intervals used for fits.
}
\label{fig:D1D2}
\end{figure}

\begin{figure}[b!]
\centering  
\includegraphics*[bb=115 365 545 715, width=0.6\textwidth]{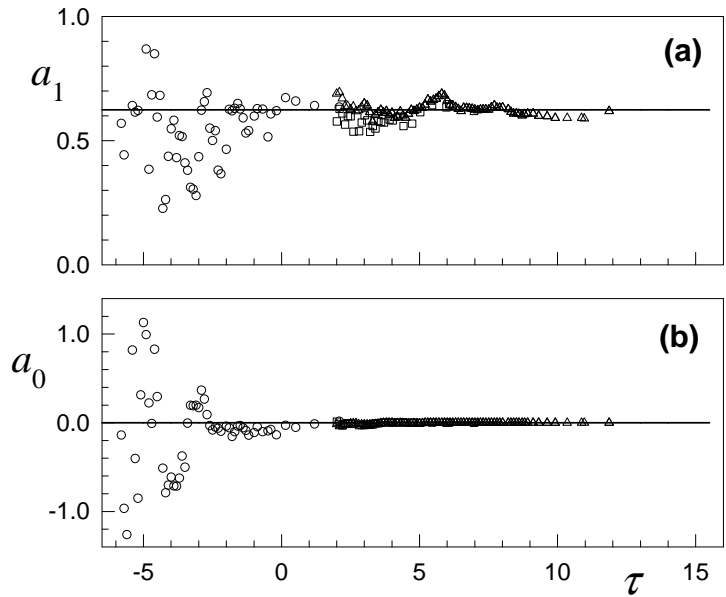} 
\caption{Dependence of drift coefficient parameters $a_1$ and $a_0$ 
on timescale $\tau$. 
Analysis was performed with daily (circles), 15-min (squares) 
and 30-sec (triangles) data series.
}
\label{fig:parsd1}
\end{figure}

\begin{figure}[ht!]
\centering  
\includegraphics*[bb=115 212 545 715, width=0.6\textwidth]{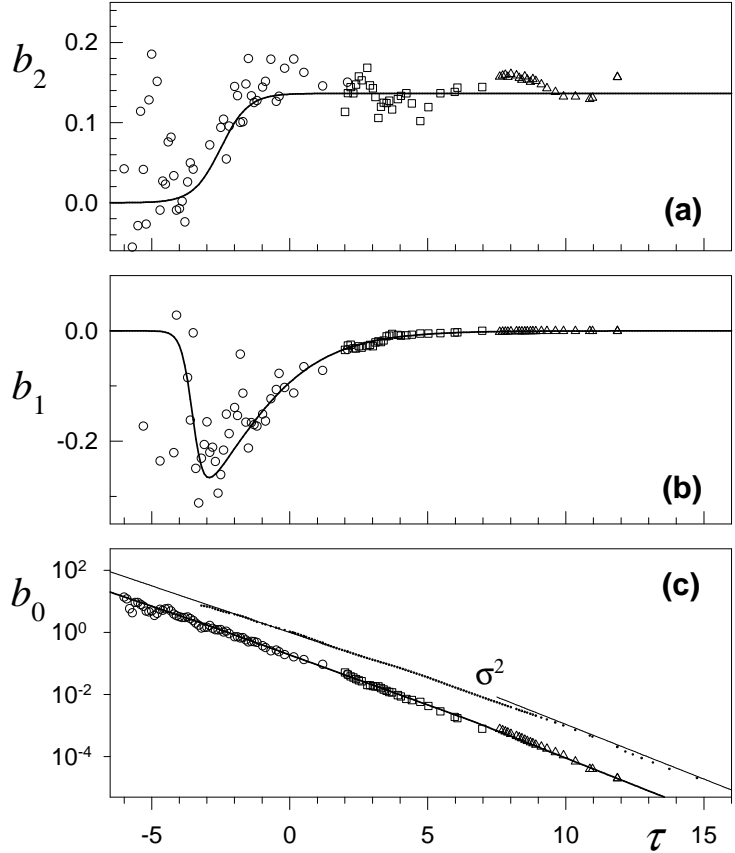}  
\caption{Dependence of diffusion coefficient 
parameters $b_2, b_1$ and $b_0$ on timescale $\tau$. Symbols as in 
Fig.~\ref{fig:parsd1}. In (c), for comparison, we include 
$\sigma^2(\tau)=\sigma^2_\tau/\sigma^2_{32}$ (small symbols) together with  
its asymptotic behaviors  (thin lines) for large and short timescales 
as predicted by Eq.~(\ref{variance}), 
with $\gamma=$1 and $1.17$, respectively.
}
\label{fig:parsd2}
\end{figure}

Markovianity was investigated by evaluating the Chapman-Kolmogorov 
equation \cite{risken}. 
We found that it holds, thus validating our approach.
The KM coefficients $D^{(k)}(\Delta x, \tau)$ 
were estimated directly from data series by means of their statistical 
definition, given by Eq.~(\ref{Dk}). 
Conditional PDFs 
$P(\Delta x_2,\tau_2 |\Delta x_1,\tau_1) = 
P(\Delta x_2,\tau_2 ;\Delta x_1,\tau_1)/P(\Delta x_1,\tau_1)$,
with $\tau_2>\tau_1$, 
were obtained from the data sets by building the histograms for the 
joint PDFs $P(\Delta x_2,\tau_2;\Delta x_1, \tau_1)$,
computed over pairs of returns $\Delta x_i$, incident at the same initial time. 
The first coefficients $\tilde D^{(k)}$ were computed for 
$\tau=(\tau_1+\tau_2)/2$ and $\Delta\tau=\tau_2-\tau_1$. 
For each couple of values ($\tau,\Delta\tau$), 
we found that  $\tilde D^{(1)}$  and $\tilde D^{(2)}$, as a function of 
$\Delta x$, follow, in very good approximation, linear and quadratic laws, respectively, 
as illustrated in Fig.~\ref{fig:D1D2}.  
Namely,
\begin{eqnarray} \label{D1}
\tilde D^{(1)} &=& -\tilde{a}_1\Delta x + \tilde{a}_0, \\ \label{D2}
\tilde D^{(2)} &=& \tilde{b}_2[\Delta x]^2 +  \tilde{b}_1\Delta x + \tilde{b}_0.  \nonumber
\end{eqnarray} 
Similar behaviors for $\tilde D^{(1)}$ and $\tilde D^{(2)}$   
have been observed for linear and logarithmic increments of 
indexes  and exchange rates involving 
US, Germany and Japan markets\cite{friedrich,amjap,peinke}. 
By fitting the linear and quadratic ansatz to the data, we obtained   
the parameters $\{\tilde{a}_i,\tilde{b}_j\}$   
for each ($\tau,\Delta\tau$).
For fixed $\tau$, the limit $\Delta\tau\to 0$ in Eq.~(\ref{Dk}) 
was achieved  by extrapolation of the parameters $\{\tilde{a}_i,\tilde{b}_j\}$ 
as a function of $\Delta \tau$. 
For the fourth order KM coefficients $\tilde{D}^{(4)}$, as a function of $\Delta x$, 
fourth order polynomial fits were performed, the limit $\Delta\tau\to 0$ 
being consistent with vanishingly small $D^{(4)}(\Delta x,\tau)$. 
Therefore, according to Pawula theorem \cite{risken} the KM expansion (\ref{KM})
can be truncated after the second order, thus reducing to the form of a FPE.
The limiting values $\{{a}_i,{b}_j\}$ determine the $\Delta x$-dependence of drift, 
$D^{(1)}$, and diffusion, $D^{(2)}$, coefficients. 
The $\tau$-dependence of the limiting parameters is presented in 
Figs.~\ref{fig:parsd1}-\ref{fig:parsd2}. 


\begin{figure}[b!]
\centering  
\includegraphics*[bb=50 210 500 800, width=0.6\textwidth]{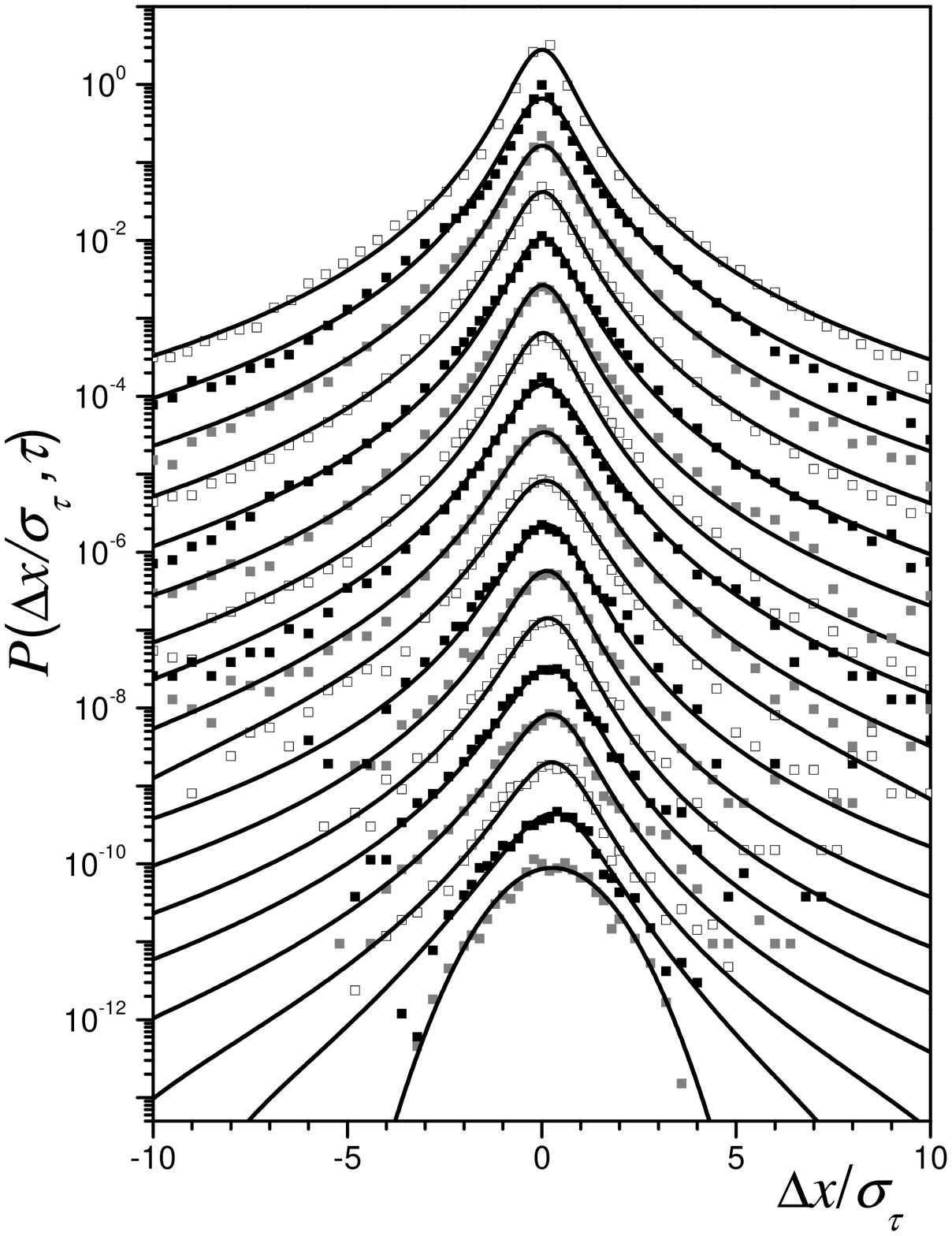} 
\caption{PDFs of normalized returns. Comparison between the numerical 
solution of  FPE (\ref{FPE}) (solid lines) and the empirical data (symbols). 
Timescales correspond to 
$\tau=-2$ to $\tau\approx 15$, from bottom to top. 
Price returns are scaled by $\sigma_\tau$ and PDFs 
shifted, for better visualization. 
A Gaussian fitted to the data for $\tau=-2$ ($\Delta t$=128 days) 
is the initial condition. 
}
\label{fig:pdfs}
\end{figure}

Parameters $a_0$ and $b_1$ describe the deviation of the PDFs from symmetry 
around zero. Indeed, this feature is observed for larger scales, 
where both parameters present non-null values, although with large fluctuations. 
In contrast, for intraday scales, $a_0$ and $b_1$ are negligible. 
Drift parameter $a_1$ remains approximately constant along daily and intraday time-scales. 
Diffusion parameter $b_2$ increases with $\tau$ from near zero up to 
a limiting value, signaling a crossover in the underlying dynamics at the monthly scale. 
Meanwhile, diffusion parameter $b_0$ presents a sustained exponential 
decay as $\tau$ increases. 
Similar behavior of $b_0$ was also reported for  
exchange rates \cite{friedrich}, although for a 
shorter range of timescales. 
Despite the exponential-like decay of $b_0$, it can 
not be neglected at large $\tau$: representing the additive noise 
component, it provides stability to the stochastic process for 
small $\Delta x$. In fact, the variance $\sigma^2_\tau$ 
also follows an exponential-like decay $2^{-\gamma\tau}$, 
as shown in Fig.~\ref{fig:parsd2}(c), 
thus setting the reference level for $b_0$.
Two limiting regimes associated with slightly different 
slopes are observed for both quantities (see Fig.~\ref{fig:parsd2}(c)),  
suggesting that $b_0(\tau)$ is related to $\sigma^2_\tau$. 
In the limit of small $\tau$ both quantities are
characterized by $\gamma=1$, corresponding to normal diffusion, 
while in the limit of large $\tau$, $\gamma>$1, 
showing the onset of  superdiffusion in the high-frequency regime. 
For $\tau>11$, data restrictions for obtaining small 
$\Delta\tau$ results prevent the estimation of the parameters.

The resulting FPE explicitly reads:
\begin{equation} \label{FPE}
\partial_\tau P =  
\partial_{\Delta x}(a_1P)+
\partial^2_{\Delta x}\left( (b_0+b_1\Delta x+b_2(\Delta x)^2)P\right),  
\end{equation}
where the $\tau$-dependence of the parameters was smoothed according to  
the ansatz plotted in Figs.~\ref{fig:parsd1}-\ref{fig:parsd2}.  
Eq.~(\ref{FPE}) was numerically integrated 
by means of a FTCS scheme \cite{recipes}.  
A Gaussian fit to the  empirical histogram at $\Delta t$=128 days 
($\tau=-2$), was used as initial condition. 
The evolution was carried out down to 
the scale of 30 seconds, the highest time resolution of our data. 
For $\tau>$11, further evolution of the FPE, 
was performed by extrapolation of the $\tau$-dependence of the coefficients. 
In Fig.~\ref{fig:pdfs}, we show the PDFs of  
returns (rescaled by $\sigma_\tau$) 
generated by 
the FPE, together with the empirical ones. 
Their agreement is remarkably good, in the full range of data,  
strongly supporting our estimation of KM coefficients.%

Within Langevin dynamics, $D^{(1)}$ and  $D^{(2)}$ 
are related to the deterministic and random forces, respectively \cite{risken}. 
Despite the wide range of analyzed time-lags, the intensity of the 
harmonic restoring force, given by  $D^{(1)}(\Delta x,\tau)\simeq-a_1(\tau)\Delta x$, 
remains almost constant for $\tau > -2$. This means that the relaxation mechanisms  
that are governed, among other factors, by constraints, flux of information, 
stock liquidity, and risk aversion, 
act similarly at diverse temporal scales.

The evolution of the diffusion coefficient presents more distinctive 
features. 
For most timescales,
$D^{(2)}(\Delta x,\tau)\simeq  b_0(\tau)+b_2(\tau)[\Delta x]^2$  is dominated by the 
state-independent and quadratic components, associated to 
additive and multiplicative noises, respectively. 
Due to the cumulative character of the fluctuations, the additive component $b_0$ 
increases with $\Delta t$. 
Meanwhile, the change of $b_2$ to a higher plateau for small $\Delta t$ 
indicates large multiplicative effects in that region, that fade away in 
the opposite limit of large time-lags, as expected.
This means that the endogenous behavior of the market, which spontaneously creates 
the amplification response mechanism to price fluctuations, presents different 
typical levels for micro and macro timescales.

The presence of multiplicative noise is known to be 
an ubiquitous mechanism to generate fat-tailed PDFs as 
steady-state solutions \cite{mult1,mult2}. 
It was found out that for a large set of control parameters, power-law tails 
prevail, with exponent depending on the 
ratio $a_1/b_2$ while being $b_0$ independent. 
Furthermore, Langevin dynamics mapped onto stochastic multiplicative 
processes with reinjection gives rise 
to power-law PDFs $P(\Delta x,\tau) \sim  |x|^{-(\mu+1)}$ \cite{mult3}, 
from which  the expression $\mu$ = $a_1/[2b_2]$ was derived \cite{sornette2}. 
From these previous results, the observed plateaux of $a_1$ and 
$b_2$  suggest a $\tau$-invariant tail of the PDF.
In fact,  the FPE (\ref{FPE}) admits invariant solutions as discussed below. 
 
Assuming an exponential law for $b_0(\tau)$, steady values of $a_1$ and $b_2$,  and neglecting 
$b_1$,  the solution of FPE (\ref{FPE}) is 
\begin{eqnarray} \label{qgaussian}
P(\Delta x,\tau) &\sim& 1/(b_0(\tau)+b_2[\Delta x]^2)^{(\mu+1)/2}\,,\mbox{    with}\\ \label{exponent}
\mu&=&1+(a_1-B/2)/b_2 \,,\\[-7mm]\nonumber
\end{eqnarray}
where $B\equiv-b'_0/b_0$  does not depend on time. 
Notice that Eq.~(\ref{exponent}) differs from the expression 
reported in \cite{sornette2}.
The solution (\ref{qgaussian}) is a $q$-Gaussian (with $q=1+2/(\mu+1)$) \cite{qgaussian}.
In the particular case $b_2\to 0$, the solutions become of the Gaussian form. 

The rescaled variance $\sigma^2(\tau)\equiv \sigma^2_\tau/\sigma^2_{32}$  is  
\begin{equation} \label{variance}
\sigma^2(\tau)=b_0(\tau)/(a_1-B/2-b_2).
\end{equation}   

In the limit of large time-lags, the diffusion coefficient is dominated by the 
state independent term $b_0$ that obeys $b_0(\tau)=b_0(0) 2^{-\tau}$. 
Substitution of the numerical values of $b_0(0)$ and $a_1$, 
into Eq.~(\ref{variance}) yields, in very good approximation, $\sigma^2(\tau)=2^{-\tau}$ 
(normal diffusion in the linear time scale) in agreement with numerical results 
(see Fig.~\ref{fig:parsd2}(c)). As a consequence, in that limit, 
the evolution equation recovers Gaussianity, ruled by a balance 
between the deterministic harmonic force and the time dependent additive noise. 
In the opposite limit of large $\tau$, $b_2$ attains a non-null steady value, while  
$b_0(\tau) \sim 2^{-\gamma\tau}$ with $\gamma>1$. 
Also in this case, Eq.~(\ref{variance}) predicts an asymptotic behavior of 
$\sigma^2$ in agreement with empirical values, as depicted in Fig.~\ref{fig:parsd2}(c). 

\begin{figure}[t!]
\centering  
\includegraphics*[bb=140 510 560 710, width=0.6\textwidth]{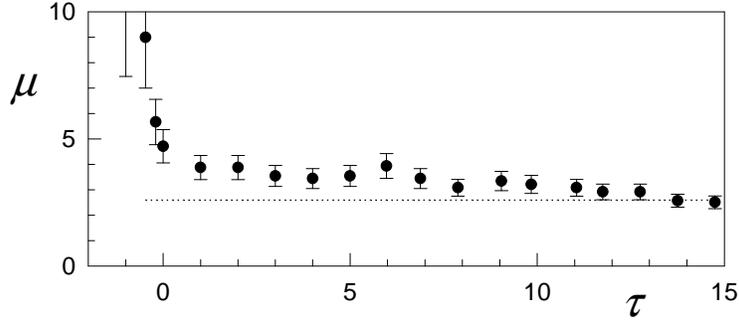}
\caption{Dependence of exponent $\mu$ on timescale $\tau$. 
Symbols correspond to fitting of Eq.~(\ref{qgaussian}) to the empirical histograms, 
the dotted line  to the asymptotic value given by Eq.~(\ref{exponent}).
}
\label{fig:nu}
\end{figure}

As the evolution of the parameters other than 
$b_0$ is slow, especially for $\tau>1$, we also checked if the PDFs could be 
effectively described by the ansatz  (\ref{qgaussian}) in an extended temporal regime. 
The outcomes of the fits 
match well the respective empirical PDFs, for almost any $\tau$. 
For $\tau\simeq 0$, the significantly non-null value of $b_1$, 
imposes a correction to the $q$-Gaussian form, yielding asymmetry. 
The values of $\mu$ resulting from (least squares) $q$-Gaussian fits 
and are displayed in Fig.~\ref{fig:nu}, together with the asymptotic value 
obtained from Eq.~(\ref{exponent}).   
For small $\tau$, the increasing value of the effective $\mu$ points 
the onset of the Gaussian regime. For large $\tau$, the empirical tail exponents 
tend to a steady value in good accord with the theoretical ones. 

Power-law tails are often quoted in the literature for financial 
assets in high-frequency regimes \cite{lisa,power-law}.
Ansatz of the $q$-Gaussian form have been 
proposed before for high-frequency \cite{lisa,next} and for 
daily \cite{ausloos} logarithmic returns from a  phenomenological approach. 
Meanwhile, in our case, they arise naturally from the evolution 
equation obtained empirically from the evaluation of KM coefficients 
throughout time-lags.

Let us recall that the results for $\tau>11$ ($\Delta t>$4 min), 
in Fig.~\ref{fig:pdfs}, were obtained 
by extrapolation of the $\tau$-dependence of the  coefficients. 
However, a good foresight of the empirical histograms arises from FPE evolution  
up to $\tau$=14.75 ($\Delta t$=1 min). 
On one hand, this ensures the reliability of the estimated parameters; on the 
other, it accounts for the predictability of intraday statistics, 
due to the well known existence of memory effects in the high-frequency regime of price returns. 
It is worth to mention that the observed deviation of the empirical histogram for the smallest 
analyzed time-lag (30 seconds) 
expresses the beginning of a non-Markovian regime \cite{small}.

In sum, we have disclosed the evolutionary pattern of the empirical 
PDFs of price returns, from Gaussian to long-tailed regimes. 
Beyond the system under study, the scope of our results may be 
appropriate for a larger financial scenario. 
Moreover our findings are not restricted to the complexity of financial data, but
may provide insights for systems on general physical contexts, such as in 
turbulent flows~\cite{swinney}, 
where similar Gaussian to $q$-Gaussian crossovers have been observed.

\acknowledgments
We acknowledge Brazilian agency CNPq for partial financial support and 
BOVESPA for providing the data.

\end{document}